\documentclass[prl,aps,showpacs,twocolumn]{revtex4}
\usepackage{epsfig}
\usepackage{bm}

\begin{document}

\title{Gradient measure and extended self-similarity of the cosmic microwave background anisotropy}

\author{\small A. Bershadskii$^{1,2}$ and K.R. Sreenivasan$^2$}
\affiliation{\small {\it $^1$ICAR, P.O. Box 31155, Jerusalem 91000, Israel}\\
{\it $^2$International Center for Theoretical Physics, Strada
Costiera 11, I-34100 Trieste, Italy}}

\begin{abstract}
Using the WMAP cosmic microwave background data it is shown that
collisions between Alfven wave packets and the cascades generated
by these collisions (the Iroshnikov model) can determine the
photon temperature fluctuations for arcminute scales on the last
scattering surface.

\end{abstract}


\maketitle

  Cosmic electromagnetic fields are expected to be generated
by the cosmological phase transitions (electroweak and QCD) in a
wide interval of scales \cite{beo}-\cite{corn} before the
recombination time. At earliest times, the magnetic fields are
generated by particle physics processes, with length scales
typical of particle physics. It is shown in \cite{beo} that
turbulence with its cascade processes is operative, and hence the
scale of magnetic fields is considerably larger than would be the
case if turbulence were ignored. In particular, it is shown in
\cite{sb},\cite{sb3},\cite{dky} that rotational velocity
perturbations, induced by a tangled magnetic field can produce
significant angular scale anisotropies in cosmic microwave
background (CMB) radiation through the Doppler effect. The
conclusions are relevant to arcminute scales \cite{sb}. For very
large cosmological scales (larger than $10^o$), on the other hand,
the Anderson localization can make it impossible for the
electromagnetic fields to propagate \cite{b}. Magnetohydrodynamic
(MHD) turbulence is characterized by a competition of two
processes, Alfven wave packets collisions and swirling motions. It
is known \cite{sb3} that on the last scattering surface the
nonlinear Alfven wave mode survives photon (Silk) damping on the
arcminute scales, while the general swirling motions (as well as
the compressional modes) are effectively dissipated on these
scales (it is significant for further consideration that the
Alfven wave modes induce specific rotational velocity
perturbations on the last scattering surface
\cite{sb},\cite{sb3},\cite{dky},\cite{adams}). Therefore, the
cosmic baryon-photon fluid becomes dominated by Alfven waves on
the arcminute scales just before the recombination time. In the
present paper we will show that specific statistical properties of
such Alfven wave dominated fluctuations are consistent with the
new (WMAP) arcminute CMB data. \\

The incompressible magnetohydrodynamic equations can be written in
terms of the Els\"{a}sser variables
$$
{\bf z}^{\pm}= {\bf v} \pm {\bf B}  \eqno{(1)}
$$
as
$$
\partial_t {\bf z}^{\pm} +{\bf z}^{\mp} \cdot {\bf \nabla} {\bf z}^{\pm}
= -{\bf \nabla} P + \nu_{+} {\bf \Delta} {\bf z}^{\pm} + \nu_{-}
{\bf \Delta} {\bf z}^{\mp},       \eqno{(2)}
$$
where $ \nu_{+}=\frac{1}{2} (\nu+\eta)$, $\nu_{-}=\frac{1}{2}
(\nu-\eta)$, $P$ is the total pressure, $\nu$ and $\eta$ are
coefficients of hydro and magnetic diffusion respectively. The
equations are given in convenient nondimensional form using units
$B/B_0 \rightarrow B$, $u/u_0 \rightarrow u$, $u_0=B_0/(4\pi
\rho)^{1/2}$ and $B_0$ is a typical magnetic field intensity.

Scaling of the structure functions of the Els\"{e}sser variables
$$
\langle |{\bf z}^{\pm} ({\bf x}+{\bf r}) -{\bf z}^{\pm} ({\bf
x})|^p \rangle \sim r^{\zeta_p}    \eqno{(3)}
$$
is used as an effective tool to study their dynamics.

A first attempt to describe magnetic turbulence dominated by
Alfven waves was made in \cite{ir}. In the incompressible fluid,
any magnetic perturbation propagates along the magnetic field
line. Since wave packets are moving along the magnetic field line,
there are two possible directions for propagation. If all the wave
packets are moving in one direction, then they are stable.
Therefore, the energy cascade occurs only when the
opposite-travelling wave packets collide, and only collisions
between similar size packets are taken into account in the
Iroshnikov model. The following amount of energy: $ \Delta E \sim
(v_l^3/l) (l/V_A) $ is released at collision of the two wave
packets of the same size $l$. The energy change per collision and
the duration of the collision are respectively $v_l^2(v_l/V_A)$
and $\Delta t \sim l/V_A$, where $V_A$ is the Alfven speed. Total
number of collisions for the cascade can be estimated as
$v^2/\Delta E$. Hence, the energy cascade time $\tau_l$ is
$$
\tau_l \sim (v^2/\Delta E)^2 \Delta t \sim \frac{l}{v_l}
\frac{V_A}{v_l}.  \eqno{(4)}
$$
That is the cascade time is $(V_A/v_l$) times longer than the eddy
turnover time $(l/v_l)$. The constancy of energy cascade:,
$(v_l^4)/(lV_A) = constant$, is assumed in the model, which
results in
$$
v_l^4 \sim l.  \eqno{(5)}
$$
It then follows from (4) and (5)
$$
\tau_l \sim l^{1/2}.   \eqno{(6)}
$$
 There exists a general representation for $\zeta_p$ (3) \cite{pp}
$$
\zeta_p={p/g} (1-x) +C_0 [1-(1-x/C_0)^{p/g}],
$$
where $g$ is related to the basic scaling $\delta v_l \sim
l^{1/g}$ (in the Iroshnikov model (5) gives $g=4$), $x$ is the
scaling of dynamic time scale of the most intermittent structures,
$\tau_l \sim l^x$ (in the Iroshnikov model (6) gives $x=1/2$),
$C_0$ is the co-dimension of these structures with the spatial
dimension $d$, $C_0=3-d$ (in the Iroshnikov model these structures
are known to be micro-sheets, i.e. $d=2$ and, consequently,
$C_0=1$). Then, for the Alfven wave dominated model \cite{gkm}
$$
\zeta_p=\frac{p}{8}+1-(1/2)^{p/4}.     \eqno{(7)}
$$
It is very significant that for this model $\zeta_4=1$ (cf. basic
scaling (5)). Using this fact, the so-called extended
self-similarity (ESS) can be introduced in the form \cite{gkm}
$$
\langle |{\bf z}^{\pm} ({\bf x}+{\bf r}) -{\bf z}^{\pm} ({\bf
x})|^p \rangle \sim \langle |{\bf z}^{\pm} ({\bf x}+{\bf r}) -{\bf
z}^{\pm} ({\bf x})|^4 \rangle^{\zeta_p}.  \eqno{(8)}
$$
The remarkable property of the ESS is that the multiscaling (8)
can survive even if the original multiscaling (3) does not exist.
Due to the local (small-scale) isotropy of the fluctuations in the
model we suppose that the ESS has the same $\zeta_p$ for any
vector component of the space difference: ${\bf z}^{\pm} ({\bf
x}+{\bf r}) -{\bf z}^{\pm} ({\bf x})$.

We use the ESS and the model (7) to check whether the cosmic
microwave background (CMB) data, obtained recently by the WMAP
space mission, support the Alfven wave domination on the arcminute
scales (cf Introduction). For this purpose we calculated moments
$\langle \Delta T_r^p \rangle$, for the space differences of the
CMB temperature fluctuations $\Delta T_r = |T({\bf x}+{\bf r}) -
T({\bf x})|$, and then checked the multiscaling
$$
\langle \Delta T_r^p \rangle \sim \langle \Delta T_r^4
\rangle^{\zeta_p}    \eqno{(9)}
$$
(cf. (8)). \\

\begin{figure}
\centering \epsfig{width=.45\textwidth,file=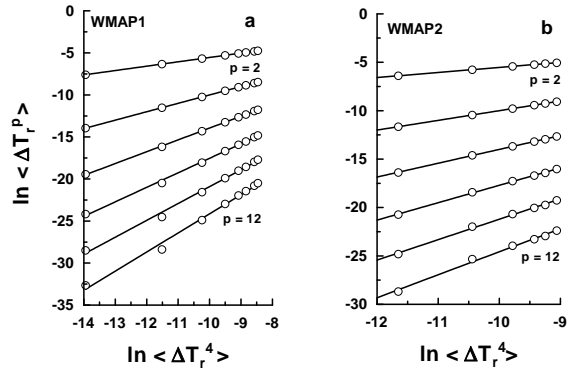}
\vspace{-6cm} \caption{{\bf a:} Logarithm of moments of different
orders $\langle|\Delta T_r|^p \rangle$ against logarithm of
$\langle|\Delta T_r|^4 \rangle$  for the cleaned and
Wiener-filtered WMAP1 data set. The straight lines (the best fit)
are drawn to indicate the scaling (9); {\bf b:} the same as in
figure 1a but for WMAP2 data set.}
\end{figure}

The motion of the scatterers imprints a temperature fluctuation,
$\delta T$, on the CMB through the Doppler effect \cite{sb},\cite
{sb3} (see also \cite{dky} and \cite{adams})
$$
\frac{\delta T ({\bf n})}{T} \sim \int g(L){\bf n} \cdot {\bf v_b}
({\bf x})~dL
$$
where ${\bf n}$ is the direction (the unit vector) on the sky,
${\bf v_b}$ is the velocity field of the baryons evaluated along
the line of sight, ${\bf x} = L{\bf n}$, and $g$ is the so-called
visibility. It should be noted that, in a potential flow, waves
perpendicular to the line of sight lack a velocity component
parallel to the line of sight; consequently, generally there is no
Doppler effect for the potential flows. The same is not true for
vortical flows, since the waves that run perpendicular to the line
of sight have velocities parallel to the line of site \cite{ov}
(see \cite{sb},\cite{sb3} about the rotational velocity
perturbations induced by the Alfven-wave modes on the last
scattering surface). One can look at the modulation from another
(though similar) point of view \cite{dky}. In general, vector
perturbations of the metric have the form
$$
\left(h_{\mu\nu}\right) =\left(\begin{array}{cc}
    0 & B_i \\
    B_j & H_{i,j} + H_{j,i}\end{array} \right),
$$
where ${\bf B}$ and ${\bf H}$ are divergence-free, 3D vector
fields supposed to vanish at infinity. The authors of \cite{dky}
introduced two gauge invariant quantities
$$
{\bf \sigma } = \dot{{\bf H}}-{\bf B}  ~~~\mbox{ and }~~~~
  {\bf z_{-}} ={\bf v-B}
$$
which represent the vector contribution to the perturbation of the
extrinsic curvature and the vorticity (cf. (6) and (13)). The
general form of the CMB temperature fluctuations produced by
vector perturbations is \cite{dky}
$$
\left(\frac{\delta T}{T}\right)^{(vec)}=
    -{\bf Z_{-}}
    \cdot {\bf n}|_{t_{dec}}^{t_0} +
  \int_{t_{dec}}^{t_0}{\bf \dot{\sigma}}\cdot {\bf n}d\lambda
$$
where ${\bf Z_{-}} = {\bf z_{-} -\sigma }$ is a gauge-invariant
generalization of the velocity field, and the subscripts {\it dec}
and $0$ denote the decoupling epoch ($z_{dec}\gg 1100$) and today
respectively. We see from this equation that, besides the Doppler
effect, Alfven waves gives rise to an integrated Sachs-Wolfe term.
However, since the geometric perturbation ${\bf \sigma}$ decays
with time, the integrated term is dominated
by its lower boundary and just cancels ${\bf \sigma}$ in $Z_{-}$. \\

 We used the WMAP data cleaned
from foreground contamination and Wiener filtered \cite{tag} (the
original temperature is given in $\mu K$ units). Wiener filtering
suppresses the noisiest modes in a map and shows the signal that
is statistically significant. However, even after the cleaning and
filtering some data points with the largest magnitudes seem to be
suspicious. Therefore, we also made several magnitude-threshold
cutoffs to check the stability of our calculations to these
cutoffs.

Figure 1a shows the results of the calculations in the form
suitable for the multiscaling of the type (9). The straight lines
in this figure (the best fits) correspond to the multiscaling (9).
The cutoff in this case excluded about 1\% of the data points (we
will call this data set WMAP1). Figure 1b shows analogous data
with a cutoff which excluded about 10\% of the data points (we
will call this data set WMAP2).
\begin{figure}
\centering \epsfig{width=.45\textwidth,file=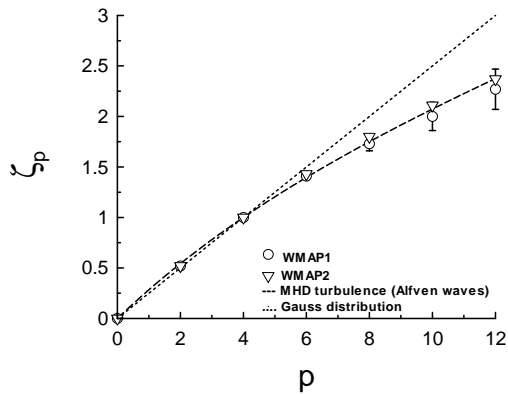}
\vspace{-5cm} \caption{The scaling exponents $\zeta_p$,
corresponding to the WMAP data (circles - WMAP1, triangles -
WMAP2). The dotted straight line corresponds to the Gaussian
distributions and the dashed curve corresponds to the Alfven-wave
dominated model calculations (7).}
\end{figure}

Figure 2 shows the exponents $\zeta_p$ (open circles correspond to
WMAP1 data set and triangles to WMAP2) extracted from figures 1a,b
as slopes of the straight lines. We also show in figure 2 as
dotted line the dependence of $\zeta_p$ on $p$ for the Gaussian
distributions ($\zeta_p = p/4$ for any Gaussian distribution), and
we show the model dependence (7) as dashed line.

One can see that starting from $p \simeq 6$ the data depart
systematically from the Gaussian straight line and follow quite
closely to the model curve (indicated by
the dashed line) predicted for the Alfven-wave dominated turbulence (7).\\

To give an additional support to the correspondence just noted
between the WMAP data and the Alfven wave dominated model, let us
introduce a gradient measure for the cosmic microwave radiation
temperature $T$ fluctuations
$$
\chi_r =\frac{\int_{v_r} (\bigtriangledown{T})^2 dv}{v_r}
\eqno{(10)},
$$
where $v_r$ is a subvolume with space-scale $r$. Scaling laws of
this measure, such as
$$
\frac{\langle \chi_{r}^p \rangle}{\langle \chi_{r} \rangle^p} \sim
r^{-\mu_p}      \eqno{(11)}
$$
are an important characteristic of the temperature dissipation
rate \cite{s}. The exponents $\mu_p$ can be related \cite{s} to
the exponents $\zeta_p$ by the equation
$$
\mu_p=1-\zeta_{4p} \eqno{(12)}
$$
that allows us to check the model equation (7) also through the
scaling (11) of the moments of the gradient measure.
\begin{figure}
\centering \epsfig{width=.45\textwidth,file=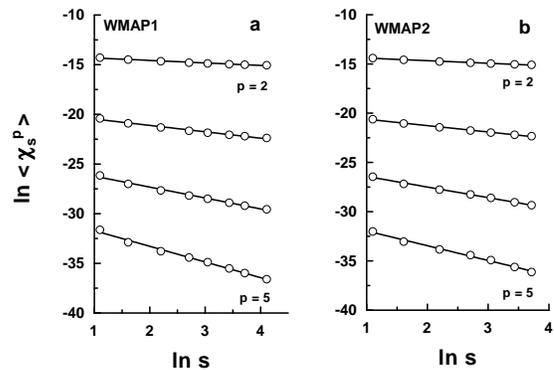}
\vspace{-5cm} \caption{{\bf a:} Logarithm of the temperature
gradient moments $\langle \chi_s^p \rangle$ calculated for the
WMAP1 data set against $\ln s$. The straight lines (the best fit)
are drawn to indicate the scaling in the log-log scales; {\bf b:}
the same as in figure 3a but for WMAP2 data set.}
\end{figure}

  Technically, using the cosmic microwave pixel data map, we will
calculate the cosmic microwave radiation temperature gradient
measure using summation over pixel sets instead of integration
over subvolumes $v_r$. The multiscaling of type (11) (if exists)
will be then written as
$$
\frac{\langle \chi_s^p \rangle}{\langle \chi_s \rangle^p} \sim
s^{-\mu_p}      \eqno{(13)}
$$
where the metric scale $r$ is replaced by number of the pixels,
$s$, characterizing the size of the summation set. The $\chi_s$ is
a surrogate of the real 3D dissipation rate $\chi_r$. It is
believed that the surrogates can reproduce quantitative
multiscaling properties of the dissipation rate \cite{s}. Since in
our calculations $\langle \chi_s \rangle$ is independent of $s$,
we will calculate the exponents $\mu_p$ directly from the scaling
of $\langle \chi_s^p \rangle$.

Figure 3a shows scaling of the CMB temperature gradient moments
$\langle \chi_s^p \rangle$ calculated for the WMAP1 map. The
straight lines (the best fit) are drawn to indicate the scaling in
log-log scales. Figure 3b shows the results of analogous
calculations produced for WMAP2 data set.

Figure 4 shows the exponents $\mu_p$ extracted from figure 3a
(circles) and from figure 3b (triangles). Dashed line in figure 4
corresponds to the model calculations (eqs. (7),(12)).\\

The results presented in the paper can be considered as a
tentative indication of the existence \cite{beo}-\cite{corn} of
the considerable magnetic fields at the recombination time. The
extended self-similarity (figure 2) shows clear non-Gaussian
character of the high moments which, together with the closeness
of the low moments to the Gaussian behavior, can shad light on the
longstanding discussion about Gaussianity of the small-scale
cosmological fluctuations. The theoretically predicted survival of
the Alfven waves on the last scattering surface at the arcminute
scales and, moreover, domination of their interactions (the
cascade collisions of the Alfven wave packets) on these scales can
be also considered now rather plausible.
\begin{figure}
\centering \epsfig{width=.45\textwidth,file=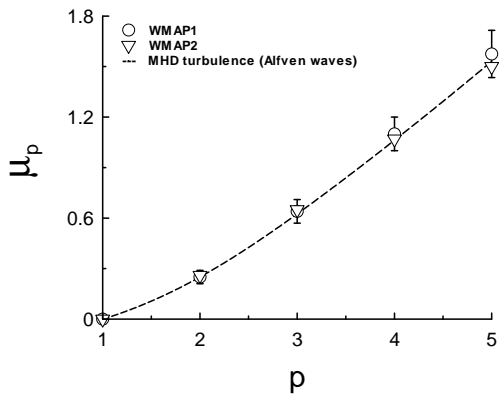}
\vspace{-5cm} \caption{ The scaling exponents $\mu_p$
corresponding to the WMAP1 data set (circles) and for the WMAP2
data set (triangles). The dashed curve corresponds to the Alfven
-wave dominated model calculations (7),(12).}
\end{figure}
WMAP sky maps produced by the WMAP team were analyzed by the team
itself. They found the CMB temperature {\it fluctuations} (i.e.
$\delta T = T- \langle T \rangle $) to obey Gaussian statistics
\cite{bennet}. Analogous situation occurs in fluid turbulence
(near-) Gaussianity of velocity fluctuations and non-Gaussian
multiscaling of the corresponding space {\it increments}
\cite{jun},\cite{ss}. The (near-) Gaussianity of a random field
fluctuations themselves does not contradict to a pronounced
non-Gaussian multiscaling of corresponding increments of the
field, which can be seen for sufficiently high order moments. This
takes place in classic fluid turbulence for velocity field and,
apparently, for the CMB temperature fluctuations (probably just
due to turbulence modulation). In particular, it was shown for
fluid turbulence \cite{jun},\cite{ss} that knowledge about the
spectrum for the Gaussian processes is certainly insufficient to
reproduce the observed multiscaling of the increments. We suppose
that there are two reasons why the previous studies of the CMB
maps failed to detect the non-Gaussianity: using the fluctuations
of the temperature (not their space increments, (9)) and the fact
that non-Gaussianity does not make its appearance in the ESS up to
the 6 moment even if space increments are considered.

    Finally, it should be noted that we cannot exclude other
contributions to high-order structure functions which may be
produced by non-linear effects due to gravity (e.g. the
Rees-Sciama effect), but the good quantitative correspondence to
turbulence (both for the structure functions and for the gradient
measure), observed in the figures 2 and 4,
indicates that the turbulence may be a dominating factor here.   \\

The authors are grateful to C.H. Gibson for discussions, and to
Tegmark's group and to the NASA Goddard Space Flight Center for
providing the data.

\end{document}